\documentclass[onecolumn,showpacs,pra,amsmath,amssymb,aps]{revtex4}
\usepackage{graphicx}
\usepackage{dcolumn}
\usepackage{bm}
\usepackage{mathrsfs}

\begin{document}

\title{Engineering Biphoton Wave Packets with an Electromagnetically Induced Grating}

\date{\today}
\author{Jianming Wen,$^{1,2}$\footnote{Email Address: jianming.wen@gmail.com} Yan-Hua Zhai,$^3$ Shengwang Du,$^4$ and Min Xiao$^{1,2}$}
\affiliation{$^1$National Laboratory of Solid State Microstructures and Department of Physics, Nanjing University, Nanjing 210093, China\\
$^2$Department of Physics, University of Arkansas, Fayetteville, Arkansas 72701, USA\\
$^3$Department of Physics, University of Maryland, Baltimore County, Baltimore, Maryland 21250, USA\\
$^4$Department of Physics, The Hong Kong University of Science and Technology, Clear Water Bay, Kowloon, Hong Kong, China}

\begin{abstract} We propose to shape biphoton wave packets with an electromagnetically induced grating in a four-level double-$\Lambda$ cold atomic system.
We show that the induced hybrid grating plays an essential role in
directing the new fields into different angular positions,
especially to the zeroth-order diffraction. A number of interesting
features appear in the shaped two-photon waveforms. For example,
broadening or narrowing the spectrum would be possible in the
proposed scheme even without the use of a cavity.
\end{abstract}

\pacs{42.50.Dv, 42.65.Lm, 42.50.Ct, 03.65.Ud}

\maketitle

\section{Introduction}
The generation of entangled paired photons with a desired joint
spectrum has become a fascinating conceptual viewpoint for both
fundamental and practical research. This is because the joint
spectrum contains the information on bandwidth, type of frequency
correlations, and wave function of the two-photon state. By
manipulating the joint spectrum, one can obtain the most appropriate
form for the specific quantum optics application under
consideration. For instance, biphotons with a narrow bandwidth play
a key role in the long-distance quantum communication protocols
based on atom-photon interface \cite{hammerer}; biphotons with a few
femtoseconds of correlation time are of particular interest in the
fields of quantum metrology \cite{giovannetti} and for some
protocols for timing and positioning measurements \cite{valencia}.

Conventionally, entangled paired photons are produced from the
process of spontaneous parametric down conversion (SPDC) in a
nonlinear crystal, where a pump photon is annihilated and two
down-converted daughter photons are simultaneously emitted
\cite{spdc}. Because of their broad bandwidth and short coherence
time, it is difficult to shape SPDC photon wave packets in the time
domain directly. A number of methods have been proposed and
developed to perform spectral manipulation of the joint spectrum
\cite{peer,viciani,hendrych} or spatial modulation of the nonlinear
interaction \cite{valencia2,harris,nasr}. Others are to modify the
(quasi-)phase matching \cite{uren}, engineer the dispersive
properties of the nonlinear medium \cite{kuzucu}, or imprint the
spectral and spatial characteristics of the pump beam into the joint
spectrum \cite{keller}.

A recent demonstration of the generation of narrow-band biphotons in
cold atomic ensembles via spontaneous four-wave mixing (SFWM)
\cite{balic,du1,du2,wen1,wen2} has attracted considerable attention
because of their long coherence time and controllable quantum wave
packets. Nonlocal modulation of temporal correlation has been
observed with such narrow-band biphotons \cite{sensarn}. In a very
recent experiment \cite{du3}, shaping of the temporal wave form by
periodically modulating the input driving lasers has confirmed the
previous theoretical prediction \cite{du4}, in which the input field
profiles can be revealed in the two-photon correlation measurements.
One major advantage over shaping the SPDC photon temporal wave
function is that these narrow-band biphotons allow further
wave-packet modification directly in the time domain.

In this paper, we describe a new way to manipulate paired Stokes and
anti-Stokes wave forms produced from SFWM in a four-level
double-$\Lambda$ \cite{xiao} cold atomic system with the use of an
electromagnetically induced grating (EIG) \cite{xiao1,araujo}. EIG
has been experimentally demonstrated in cold atoms
\cite{imoto,cardoso} and has been applied to all optical switching
and routing in hot atomic vapors \cite{xiao2}. Here, we show that,
by spatially modulating the control beams, alternating regions of
high transmission and absorption can be created inside the atomic
sample that act as an amplitude grating and by which the joint
Stokes and anti-Stokes wave packet can be shaped. Compared with
previous proposals ascribed above, several interesting features
appear in the present one. First, such a medium may exert both
amplitude and phase modulations on biphoton wave packets in much the
same way that a hybrid (amplitude and phase) grating does to the
amplitude and phase of an electromagnetic wave. Second, the spatial
modulation of the control fields is imprinted into both the linear
and the nonlinear susceptibilities. Consequently, this mapping may
broaden or narrow the joint spectrum depending on the system's
parameters. Third, but not least, because of the grating diffraction
interference, the spectral brightness can be improved, and the
emission angle can be confined to some particular angles. For
example, the anti-Stokes field will be mainly directed to the
zeroth-order diffraction.

We organize the paper as follows. The basic idea is presented in
Sec. II by considering two-photon temporal correlation measurement.
The conclusion is summarized in Sec. III.

\section{Shaping Biphoton Wave Form with EIG}

\subsection{EIG}
To illustrate the basic idea, we consider a four-level
double-$\Lambda$ atomic system (e.g. $^{87}$Rb) depicted in
Fig.~1(a), where all the atomic population is assumed to be in the
ground state $|1\rangle$. To ignore the Doppler broadening, the
atoms are laser cooled in a magnetic optical trap. Two strong
control fields ($\omega_c$), resonant with the atomic transition
$|2\rangle\rightarrow|3\rangle$ while being symmetrically displaced
with respect to $z$, are incident upon the atomic ensemble at such
angles that they intersect and form a standing wave within the
medium [see Fig. 1(c)]. In the presence of the counterpropagating
weak probe field ($\omega_p$) far detuned from the transition
$|1\rangle\rightarrow|4\rangle$, phase matched Stokes ($\omega_s$)
and anti-Stokes ($\omega_{as}$) photons are then spontaneously
generated in opposite directions and are detected by single-photon
detectors D$_2$ and D$_1$, respectively, as shown in Fig. 1(b).
Since the linear and nonlinear optical responses to the generated
fields depend on the strength of the control light, they are
expected to change periodically as the standing wave changes from
the nodes to anti-nodes across the $x$ dimension. In the current
configuration, the Stokes photons travel at nearly the speed of
light in vacuum with negligible Raman gain. In contrast, the strong
control beams induce a set of periodic transparency windows to the
anti-Stokes field. Thus, alternatively, a nonmaterial grating is
formed in the anti-Stokes channel. This grating is termed as EIG
\cite{xiao1}, which will diffract the anti-Stokes field into some
particular angles according to the diffraction orders.

\begin{figure}[tbp]
\includegraphics[scale=0.5]{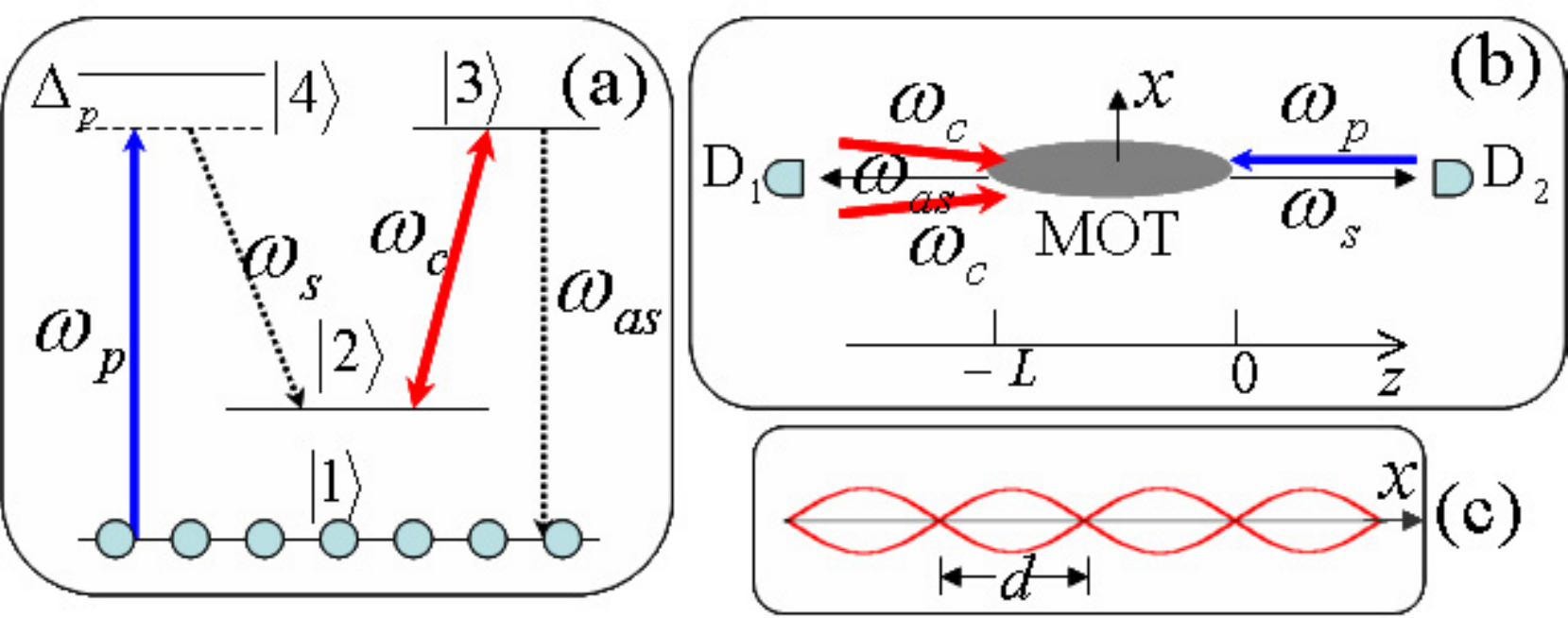}
\caption{(Color online) Shaping biphoton wave packets with an EIG.
(a) The level structure, where in the presence of a cw probe
($\omega_p$) and control ($\omega_c$) fields, paired Stokes
($\omega_s$) and anti-Stokes ($\omega_{as}$) photons are
spontaneously created from the four-wave mixing processes in the
low-gain regime. (b) The backward generation geometry, where two
strong control beams symmetrically displace with respect to $z$ and
form a standing wave along $x$. (c) The standing wave formed by
control fields.}\label{fig:Fig1}
\end{figure}

Following the analysis presented in Ref. \cite{wen1}, the
third-order nonlinear susceptibility for the generated anti-Stokes
field is calculated to be
\begin{eqnarray}
\chi^{(3)}_{as}(\omega)
=\frac{-N\mu_{13}\mu_{32}\mu_{24}\mu_{41}/[4\hbar^3\epsilon_0(\Delta_p+i\gamma_{41})]}{(\omega-\Omega_e+i\gamma_e)(\omega+\Omega_e+i\gamma_e)},
\end{eqnarray}
and the linear susceptibilities at the Stokes and anti-Stokes
frequencies are, respectively,
\begin{eqnarray}
\chi_s(\omega)&=&\frac{N|\mu_{42}|^2(\omega-i\gamma_{31})/(4\hbar\epsilon_0)}{|\Omega_c|^2\cos^2
(\frac{\pi{x}}{d})-(\omega-i\gamma_{31})(\omega-i\gamma_{21})}\frac{|\Omega_p|^2}{\Delta^2_p+\gamma^2_{41}},\nonumber\\
\chi_{as}(\omega)&=&\frac{N|\mu_{31}|^2(\omega+i\gamma_{21})/(\hbar\epsilon_0)}{|\Omega_c|^2\cos^2(\frac{\pi{x}}{d})-(\omega+i\gamma_{31})(\omega+i\gamma_{21})},
\end{eqnarray}
where $N$ is the atomic density, $\mu_{ij}$ are dipole matrix
elements, $\Omega_p$ and $\Omega_c$ are the probe/control Rabi
frequency, $\gamma_{ij}$ are the decay or dephasing rate,
$\Delta_p=\omega_p-\omega_{41}$ is the probe detuning, and
$d=\frac{\pi}{k_{cx}}$ represents the space period, which can be
made arbitrarily larger than the wavelength of the control fields by
varying the angle between their two wave vectors.
$\Omega_e=\sqrt{|\Omega_c|^2\cos^2(\frac{\pi{x}}{d})+\gamma_{31}\gamma_{21}}\approx|\Omega_c|\cos(\frac{\pi{x}}{d})$
is the effective control Rabi frequency, and
$\gamma_e=\frac{\gamma_{31}+\gamma_{21}}{2}$ is the effective
dephasing rate. $\chi^{(3)}_{as}$ in Eq.~(1) has two resonances
separated by $\Omega_e$ and each is associated with a linewidth of
$2\gamma_e$. From Eqs.~(1) and (2), it is obvious that the spatial
periodic modulation of the control fields has been mapped into the
optical responses to the Stokes and anti-Stokes fields.
Consequently, such a modulation will further modify the two-photon
wave form as will be discussed later. It is known that the linear
susceptibilities determine the transmission bandwidth and dispersion
property. Taking $|\Omega_p|\ll\Delta_p$, $\chi_s$ is approximated
as 0, which means the Stokes photons traverse the medium almost at
the speed of light in vacuum, and the Raman gain is negligible. In
contrast, the anti-Stokes photons may propagate at a lower group
velocity,
$v_g\approx2\hbar\epsilon_0c|\Omega_c|^2\cos^2(\frac{\pi{x}}{d})/N|\mu_{31}|^2\omega_{31}=v_0\cos^2(\frac{\pi{x}}{d})$,
and experience periodic linear loss characterized by
$\alpha=N\sigma_{31}\gamma_{21}\gamma_{31}/2[|\Omega_c|^2\cos^2(\frac{\pi{x}}{d})+\gamma_{21}\gamma_{31}]$,
where
$\sigma_{31}=\omega_{31}|\mu_{31}|^2/(\hbar\epsilon_0c\gamma_{31})$
is the on-resonance absorption cross section in the transition
$|1\rangle\rightarrow|3\rangle$.

Thus, such a periodic linear loss results in an EIG to the
anti-Stokes photons. Figure 2 displays a typical transmission
function for the anti-Stokes light as a function of $x$. It is easy
to understand that, at the transverse locations around the nodes (of
the standing wave), the control field intensities are so weak that
the anti-Stokes field is absorbed according to the usual Beer law.
In contrast, since the intensity distribution of the control fields
at the spatial locations around the antinodes is very strong, the
absorption of the anti-Stokes field is greatly suppressed due to the
effect of electromagnetically induced transparency \cite{EIT}. This
leads to a periodic amplitude modulation across the beam profile of
the anti-Stokes light, a phenomenon reminiscent of the amplitude
grating.

\begin{figure}[tbp]
\includegraphics[scale=0.35]{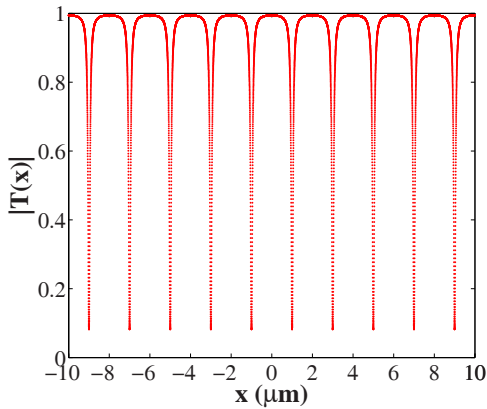}
\caption{A typical transmission profile of the anti-Stokes field as
a function of $x$. Parameters are chosen as $d=2$ $\mu$m, the
optical depth about 5, $\gamma_{31}=2\pi\times3$ MHz, and
$\gamma_{21}=0.6\times\gamma_{31}$.}\label{fig:Fig2}
\end{figure}

\subsection{Shaping two-photon wave form}
The paired Stokes and anti-Stokes photon state can be obtained from
first-order perturbation theory \cite{du2,wen1,rubin}. For
simplicity, we take the input probe and control beams as classical
cw lasers and focus on the two-photon temporal correlation. The
effective interaction length is taken as $L$. The unnormalized
biphoton state at the output surfaces of the sample may be written
as
\begin{eqnarray}
|\Psi\rangle=\Psi_0\int{d}\omega\Phi(\omega)a^{\dag}_sa^{\dag}_{as}|0\rangle,
\end{eqnarray}
where $\Psi_0$ is a grouped constant, and the joint spectral
function takes the form
\begin{eqnarray}
\Phi(\omega)=\int^{x_2}_{x_1}dx\chi^{(3)}_{as}(\omega)\cos\bigg(\frac{\pi{x}}{d}\bigg)\mathrm{sinc}\bigg(\frac{\Delta{k}L}{2}\bigg)e^{i\frac{\Delta{k}L}{2}},
\end{eqnarray}
where the cosine term comes from the standing wave of the control
fields, and the last two terms from the longitudinal phase matching
condition with $\Delta{k}\approx\frac{\omega}{v_g}+i\alpha$. In Eq.
(4), we have taken the linear loss into account. It is clear that
that the joint spectrum $\Phi$ can be engineered through
$\chi^{(3)}_{as}$ and the phase matching condition. After, we
describe the wave-packet shaping by considering the simple
two-photon temporal correlation measurement in which paired Stokes
and anti-Stokes photons are detected by single-photon detectors
D$_1$ and D$_2$ with equal pathways from the output surfaces of the
medium, as shown in Fig. 1(b). Since there are two characteristic
timings embedded in Eq. (4), the resonance linewidth determined by
$\chi^{(3)}_{as}$ and the natural spectral width determined by the
phase matching condition $\frac{L}{v_0}$ will be looked at
separately by using the two-photon temporal correlation in which
only one characteristic timing is dominant.

Using the Glauber theory, the two-photon amplitude is
\begin{eqnarray}
A=\langle0|E^{(+)}_sE^{(+)}_{as}|\Psi\rangle.
\end{eqnarray}
The field $E^{(+)}_j$ is the positive-frequency part of the
free-space electromagnetic field at position $r_j$ and time $t_j$.
In the far-field region (Fraunhofer diffraction), the biphoton
amplitude (5) over the diffraction angle $\theta$ (with respect to
$z$) can be derived, by following the procedure done in Refs.
\cite{wen1,wen2,wen3}, as
\begin{eqnarray}
A(\tau;\theta)=A\int^{x_2}_{x_1}dx\cos\bigg(\frac{\pi{x}}{d}\bigg)e^{ik_{as}x\sin\theta}
\int{d}\omega\chi^{(3)}_{as}(\omega)\mathrm{sinc}\bigg(\frac{\Delta{k}L}{2}\bigg)e^{i(\frac{\Delta{k}L}{2}-\omega\tau)},
\end{eqnarray}
where $A$ is an integrant-irrelevant constant, $\tau=t_{as}-t_s$ is
the relative time delay between two clicks, and $k_{as}$ is the
central wave number of the anti-Stokes photons. Equation (6) can
further be recast into a product of an integral and a geometric
series
\begin{eqnarray}
A(\tau;\theta)=A\sum^{M/2}_{n=-M/2}e^{ik_{as}nd\sin\theta}\int^{\frac{d}{2}}_{-\frac{d}{2}}dx\cos\bigg(\frac{\pi{x}}{d}\bigg)
{e}^{ik_{as}x\sin\theta}\int{d}\omega\chi^{(3)}_{as}(\omega)\mathrm{sinc}\bigg(\frac{\Delta{k}L}{2}\bigg)e^{i(\frac{\Delta{k}L}{2}-\omega\tau)},
\end{eqnarray}
where $M$ represents the input probe field across $M$ times $d$.
This can be guaranteed by adjusting the diameters of both probe and
control fields to cover $M$ slits. By evaluating the geometric
progression in the usual fashion, Eq. (7) can be written as
\begin{eqnarray}
A(\tau;\theta)=A\frac{\sin\frac{k_{as}Md\sin\theta}{2}}{\sin\frac{k_{as}d\sin\theta}{2}}B(\tau;\theta),
\end{eqnarray}
with
\begin{eqnarray}
B(\tau;\theta)=\int^{\frac{d}{2}}_{-\frac{d}{2}}dx\cos\bigg(\frac{\pi{x}}{d}\bigg)e^{ik_{as}x\sin\theta}
\int{d}\omega\chi^{(3)}_{as}(\omega)\mathrm{sinc}\bigg(\frac{\Delta{k}L}{2}\bigg)e^{i(\frac{\Delta{k}L}{2}-\omega\tau)}.
\end{eqnarray}
Therefore, the diffracted two-photon amplitude is a product of a
single \textit{slit} Eq. (9) multiplied by the function in Eq. (8).
Equations (8) and (9) together imply that the two correlated Stokes
and anti-Stokes photons are simultaneously produced from any one of
the \textit{slits}, which can be regarded as a superposition of
coherent SFWM subsources. We also notice that the first integration
in Eq. (9) can be visualized as an amplitude grating with a
transmission profile followed by a cosine curvature. The emission
angles and diffraction efficiencies are determined by the ability of
the induced grating. From Eq. (8), it is easy to obtain the
diffraction angles of the anti-Stokes field for different
diffraction orders $m$ as
\begin{eqnarray}
\sin\theta=m\frac{\lambda_{as}}{d},
\end{eqnarray}
where $\lambda_{as}=2\pi/k_{as}$. According to the results shown in
Ref. \cite{xiao1}, the diffraction mainly occurs at the zeroth
order. This could be important to direct the light into a smaller
solid angle and, hence, enhance its spectral brightness at the
observation's location. Equations (8) and (9) are our starting
points to analyze shaping of biphoton wave forms using EIG. Since,
for the anti-Stokes field, the energy is almost emitted toward the
zeroth-order diffraction direction, we assume $\theta=0$ in Eq. (8)
to simplify the analysis in the following.

\subsection{Two-photon coincidence counts}
First, let us look at the case in which the coherence time is mainly
determined by the resonance linewidth. In such a case, the natural
spectral width from the phase matching is much greater than the
linewidth. Hence, its effect on two-photon temporal correlation can
be ignored. Thus, Eq. (9) reduces to
\begin{eqnarray}
B(\tau)=\int^{\frac{d}{2}}_{-\frac{d}{2}}dx\cos\bigg(\frac{\pi{x}}{d}\bigg)\int{d}\omega\chi^{(3)}_{as}(\omega)e^{-i\omega\tau}.
\end{eqnarray}
Plugging Eq. (1) into Eq. (11) and completing the frequency integral
yields
\begin{eqnarray}
B(\tau)=B\int^{\frac{d}{2}}_{-\frac{d}{2}}dx\sin\bigg[|\Omega_c|\tau\cos\bigg(\frac{\pi{x}}{d}\bigg)\bigg]e^{-\gamma_e\tau},
\end{eqnarray}
where $B$ is a constant. Different from previous findings
\cite{balic,du1,du2,wen1,wen2,du3,du4,wen3}, Eq. (12) clearly shows
that the profile of the biphoton wave form is further manipulated by
the periodic modulation of the control fields. Implementing the
integration in Eq. (12) gives
\begin{eqnarray}
B(\tau)=BdH_0(|\Omega_c|\tau)e^{-\gamma_e\tau},
\end{eqnarray}
where $H_0(x)$ is the Struve function of order zero. The two-photon
coincidence counting rate equals the square of $A(\tau)$, whose
profile is governed by $H_0(x)$ and is manifested by an exponential
decay. In Fig. 3, we have provided two typical simulations of the
coincidences using the parameters in Ref.~\cite{du2}. We notice that
the damped oscillations shown in Fig. 3 do not obey Rabi flopping as
previously reported in Refs. \cite{balic,du1,du2,wen2,wen3}. The
origin of this difference comes from the periodic modulation of the
control fields, which, in turn, modifies the joint-detection
patterns. The minimum coincidences appear at the zero solutions of
$H_0(x)$. The lower curve in Fig. 3 gives the over-damped case in
which even a single oscillation is not fully observable because of
the fast exponential decay. Another noticeable feature is that the
joint spectrum is broadened in a single oscillation due to the
diffraction interference.

\begin{figure}[tbp]
\includegraphics[scale=0.42]{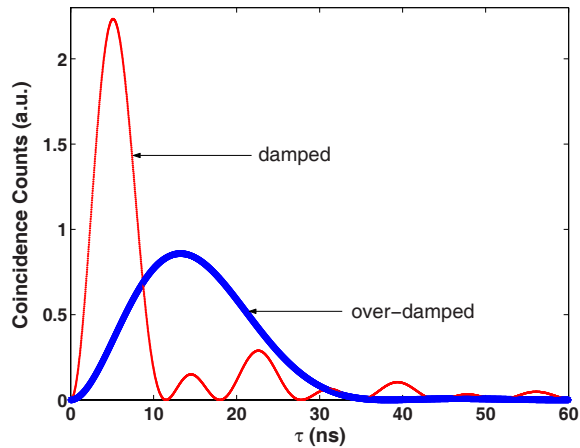}
\caption{(Color online) Two-photon temporal coincidences exhibit
damped and overdamped oscillations with the space period $d=2$
$\mu$m. Other parameters are the same as in Ref.
\cite{du2}.}\label{fig:Fig3}
\end{figure}

Next, we look at the two-photon temporal correlation mainly
characterized by the phase-matching condition. That is, the natural
spectral width is much narrower than the resonance linewidth such
that the intrinsic mechanism of biphoton generation is partially or
even fully washed out. (The latter case requires much higher optical
depth.) In such a case, Eq. (9) becomes
\begin{eqnarray}
B(\tau)=\int^{\frac{d}{2}}_{-\frac{d}{2}}dx\cos\bigg(\frac{\pi{x}}{d}\bigg)
\int{d}\omega\mathrm{sinc}\bigg(\frac{\Delta{k}L}{2}\bigg)e^{i(\frac{\Delta{k}L}{2}-\omega\tau)},
\end{eqnarray}
which can be numerically evaluated. In Fig. 4, we have plotted the
coincidence counting rate with the space period $d=2$ $\mu$m plus
taking the third-order nonlinearity [Eq. (9)] into account. As
illustrated in Fig. 4(a), most of the features appearing in previous
studies [for instance, see Fig. 4(b)] can be observed. For example,
the sharp peak in the leading edge of the two-photon coincidence
counts represents the Sommerfeld-Brillouin precursor at the biphoton
level, as report in Ref. \cite{du5}. One difference from previous
results in the literature \cite{du1,du2,du4} is that, at the tail in
Fig. 4(a), several small bumps emerge instead of a smooth
exponential decay. Another difference is that the coherence time is
extended. In Fig. 4(a), the coherence time is extended by more than
1 $\mu$s. However, without the induced grating as shown in Fig.
4(b), the coherence time is only about 800 ns. Alternatively, the
joint spectrum of biphotons is narrowed. This spectrum narrowing is
a result of the spatial modulation of the control fields plus the
modulated group velocities of the anti-Stokes field. Without the use
of the cavity, the spectrum narrowing achieved here is useful for
producing narrow-band biphotons with higher spectral brightness. If
the optical depth of the medium could be made enough high, the
two-photon temporal correlation would be closer to a square-wave
pattern as usually observed in the SPDC process. Those small bumps
would become discrete step functions at the tail, which can be
verified from Eq. (14). Since this looks more like an ideal case and
might not be detectable in the experiment, we will not offer further
discussions here.

\begin{figure}[tbp]
\includegraphics[scale=0.42]{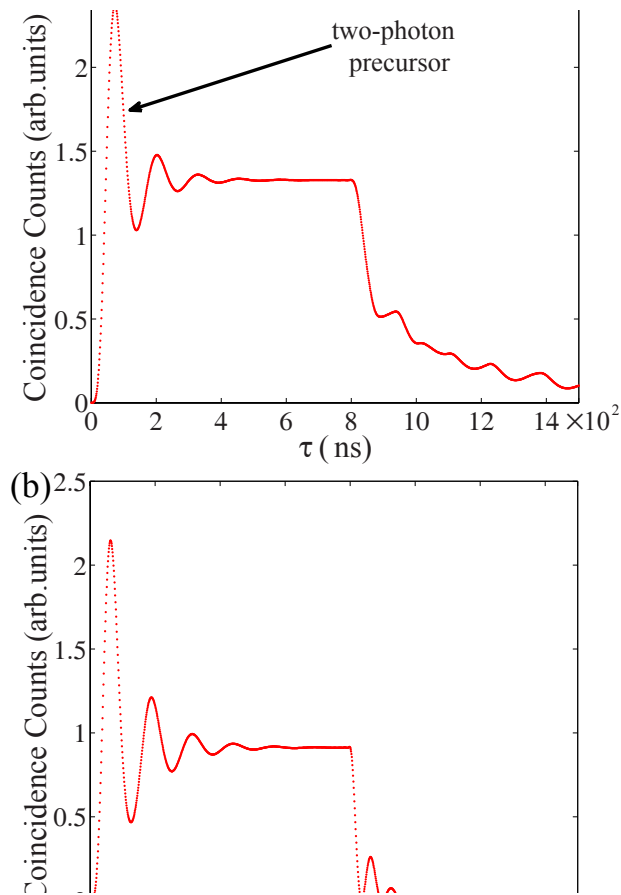}
\caption{(Color online) Two-photon temporal coincidences: (a)
modulated by an EIG with the space period $d=2$ $\mu$m. Other
parameters are chosen as $L/v_0=800$ ns, $\Omega_c=5\gamma_{31}$,
$\gamma_{31}=2\pi\times3$ MHz, and $\gamma_{21}=0.6\gamma_{31}$. (b)
without the induced grating. Same parameters are chosen as in
(a).}\label{fig:Fig4}
\end{figure}

Before ending the discussions, in Secs. IIA and IIB, we have
analyzed how to shape the entangled Stokes-anti-Stokes temporal wave
form with the use of EIG. The extension of the idea to be used on a
nonlinear crystal would be interesting. Although it is easy to
design a diffraction grating within or at the output surface of the
crystal, it is difficult to modulate the dispersion periodically and
spatially. Therefore, it is very challenging to fully recover the
features obtained here in nonlinear crystals.

\section{Summary}
In summary, here, we have proposed a method to engineer the
two-photon temporal wave packets by utilizing an EIG. The method
distinguishes itself from previous research by the appearance of
several features. First, the induced grating influences both the
linear and the nonlinear susceptibilities. As a consequence, this
will shape the biphoton wave packets through both the dispersive
properties of the medium and the periodic nonlinear optical
responses. Second, the induced (hybrid) nonmaterial grating directs
the output anti-Stokes field into different angular positions,
especially into its zeroth-order diffraction. Third, the modulated
biphoton wave packets exhibit different profiles compared with
previous studies. For example, the damped oscillations do not
coincide with the Rabi oscillations as observed in Refs.
\cite{balic,du1,du2}. The decayed square-wave pattern shows small
bumps at the tail, which, to the best of our knowledge, have never
been discovered in the literature. Fourth, the spectral brightness
and emission angle can be further engineered by the induced grating.
This paper is important not only because it explores another
application of the EIG, but also because the shaped biphoton wave
pakckets hold applications in certain protocols of quantum
information, quantum communications, and quantum cryptography. For
instance, the properties ascribed before can be used to direct the
propagation of single photons and improve the efficiency of
detecting photons in free space due to the diffraction. The
broadened or narrowed bandwidth could be useful for coherent
absorption and reemission of photons based on the interface between
atoms and photons. The effect of EIG on transverse correlation of
entangled photons may be interesting and worth studying. However,
such an issue is beyond the scope of the current paper and might be
addressed somewhere else.

\section{Acknowledgements}

We gratefully acknowledge insightful discussions with M. H. Rubin,
K.-H. Luo, and Xiaoshun Jiang. J.W. and M.X. were supported, in
part, by the National Science Foundation (USA). J.W. also
acknowledges financial support by 111 Project No. B07026 (China).
S.D. was supported by the Hong Kong Research Grants Council (Project
No. HKUST600809).


\begin{thebibliography}{99}

\bibitem{hammerer} K. Hammerer, A. S. S\o{r}ensen, and E. Polzik, Rev. Mod. Phys. \textbf{82}, 1041 (2010).

\bibitem{giovannetti} V. Giovannetti, S. Lloyd, and L. Maccone, Phys. Rev. Lett. \textbf{96}, 010401 (2006).

\bibitem{valencia} A. Valencia, G. Scarcelli, and Y.-H. Shih, Appl. Phys. Lett. \textbf{85}, 2655 (2004).

\bibitem{spdc} Y.-H. Shih, Rep. Prog. Phys. \textbf{66}, 1009 (2003).

\bibitem{peer} A. Pe'er, B. Dayan, A. A. Friesem, and Y. Silberberg, Phys. Rev. Lett. \textbf{94}, 073601 (2005).

\bibitem{viciani} M. Bellini, F. Marin, S. Viciani, A. Zavatta, and F. T. Arecchi, Phys. Rev. Lett. \textbf{90}, 043602 (2003).

\bibitem{hendrych} M. Hendrych, X. Shi, A. Valencia, and J. P. Torres, Phys. Rev. A \textbf{79}, 023817 (2009).

\bibitem{valencia2} A. Valencia, A. Cer\'{e}, X. Shi, G. Molina-Terriza, and J. P. Torres, Phys. Rev. Lett. \textbf{99}, 243601 (2007).

\bibitem{harris} S. E. Harris, Phys. Rev. A \textbf{78}, 021807(R) (2008).

\bibitem{nasr} M. B. Nasr, S. Carrasco, B. E. A. Saleh, A. V. Sergienko, M. C. Teich, J. P. Torres, L. Torner, D. S. Hum,
and M. M. Fejer, Phys. Rev. Lett. \textbf{100}, 183601 (2008).

\bibitem{uren} A. B. U'Ren, R. K. Erdmann, M. de la Cruz-Gutierrez, and I. A. Walmsley, Phys. Rev. Lett. \textbf{97}, 223602 (2006).

\bibitem{kuzucu} O. Kuzucu, M. Fiorentino, M. A. Albota, F. N. C. Wong, and F. X. K\"{a}rtner, Phys. Rev. Lett. \textbf{94}, 083601 (2005).

\bibitem{keller} T. E. Keller and M. H. Rubin, Phys. Rev. A \textbf{56}, 1534 (1997).

\bibitem{balic} V. Bali\'{c}, D. A. Braje, P. Kolchin, G. Y. Yin, and S. E. Harris, Phys. Rev. Lett. \textbf{94}, 183601 (2005).

\bibitem{du1} S. Du, P. Kolchin, C. Belthangady, G. Y. Yin, and S. E. Harris, Phys. Rev. Lett. \textbf{100}, 183603 (2008).

\bibitem{du2} S. Du, J.-M. Wen, and M. H. Rubin, J. Opt. Soc. Am. B \textbf{25}, C98 (2008).

\bibitem{wen1} J.-M. Wen and M. H. Rubin, Phys. Rev. A \textbf{74}, 023808 (2006); \textbf{74}, 023809 (2006).

\bibitem{wen2} J.-M. Wen, S. Du, and M. H. Rubin, Phys. Rev. A \textbf{75}, 033809 (2007);
S. Du, J.-M. Wen, M. H. Rubin, and G. Y. Yin, Phys. Rev. Lett.
\textbf{98}, 053601 (2007).

\bibitem{sensarn} S. Sensarn, G. Y. Yin, and S. E. Harris, Phys. Rev. Lett. \textbf{103}, 163601 (2009).

\bibitem{du3} J. F. Chen, S. Zhang, H. Yan, M. M. T. Loy, G. K. L. Wong, and S. Du, Phys. Rev. Lett. \textbf{104}, 183604 (2010).

\bibitem{du4} S. Du, J.-M. Wen, and C. Belthangady, Phys. Rev. A \textbf{79}, 043811 (2009).

\bibitem{xiao} B. L. L\"{u}, W. H. Burkett, and M. Xiao, Opt. Lett.
\textbf{23}, 804 (1998).

\bibitem{xiao1} H. Y. Ling, Y.-Q. Li, and M. Xiao, Phys. Rev. A \textbf{57}, 1338 (1998).

\bibitem{araujo} L. E. E. de Araujo, Opt. Lett. \textbf{35}, 977 (2010).

\bibitem{imoto} M. Mitsunaga and N. Imoto, Phys. Rev. A \textbf{59}, 4773 (1999).

\bibitem{cardoso} G. C. Cardoso and J. W. R. Tabosa, Phys. Rev. A \textbf{65}, 033803 (2002).

\bibitem{xiao2} A. W. Brown and M. Xiao, Opt. Lett. \textbf{30}, 699 (2005).

\bibitem{EIT} See, for examples, S. E. Harris, Phys. Today 50 (7), 36 (1997); M. Xiao,
Y.-q. Li, S. Z. Jin, and J. Gea-Banacloche, Phys. Rev. Lett.
\textbf{74}, 666 (1995).

\bibitem{rubin} M. H. Rubin, D. N. Klyshko, Y.-H. Shih, and A. V. Sergienko, Phys. Rev. A \textbf{50}, 5122 (1994).

\bibitem{wen3} J.-M. Wen, S. Du, Y. P. Zhang, M. Xiao, and M. H. Rubin, Phys. Rev. A \textbf{77}, 033816 (2008).

\bibitem{du5} S. Du, C. Belthangady, P. Kolchin, G. Y. Yin, and S.
E. Harris, Opt. Lett. \textbf{33}, 2149 (2007).

\end{thebibliography}
\end{document}